\documentclass[twocolumn]{aastex631}

\usepackage{comment}

\shorttitle{Galactic-scale magnetic winds from NGC 1532}
\shortauthors{Matthews et al.}
\graphicspath{{./}{figures/}}

\begin{document}

\title{A Galactic Scale Magnetized Wind Around a Normal Star-Forming Galaxy}

\author[0000-0002-6479-6242]{A.~M.~Matthews}
\affiliation{Observatories of the Carnegie Institution for Science,
813 Santa Barbara St,
Pasadena, CA 91101, USA}

\author[0000-0001-7363-6489]{W.~D.~Cotton}
\affiliation{National Radio Astronomy Observatory,
520 Edgemont Road,
Charlottesville, VA 22903, USA}

\author[0000-0002-5187-7107]{W.~M.~Peters}
\affiliation{U.S. Naval Research Laboratory, 4555 Overlook Ave. SW, Washington, DC 20375, USA}

\author[0000-0003-3948-7621]{L.~Marchetti}
\affiliation{Department of Astronomy, University of Cape Town, 7701 Rondebosch, Cape Town, South Africa}
\affiliation{INAF - Instituto di Radioastronomia, via Gobetti 101, I-40129 Bologna, Italy}
\affiliation{The Inter-University Institute for Data Intensive Astronomy (IDIA), Department of Astronomy, University of Cape Town, 7701 Rondebosch, Cape Town, South Africa}

\author[0000-0002-4939-734X]{T.~H.~Jarrett}
\affiliation{Department of Astronomy, University of Cape Town, 7701 Rondebosch, Cape Town, South Africa}

\author[0000-0003-4724-1939]{J.~J.~Condon}
\affiliation{Independent Researcher}

\author[0000-0000-0000-0000]{J.~M.~van der Hulst}
\affiliation{Kapetyn Astronomical Institute, University of Groningen, Landleven 12, 9747AD Groningen, The Netherlands}

\author[0000-0001-5519-0620]{M.~Moloko}
\affiliation{Department of Astronomy, University of Cape Town, Rondebosch, 7700, South Africa}


\begin{abstract}

Galaxy formation theory identifies superwinds as a key regulator of star formation rates, galaxy growth, and chemical enrichment. Thermal and radiation pressure are known to drive galactic-scale winds in dusty starbursting galaxies (e.g. M82), but modern numerical simulations have recently highlighted that cosmic-ray (CR) driven winds may be especially important in normal galaxies with modest star formation rate surface densities. However, CR-driven winds have yet to be conclusively observed---leaving significant uncertainty in their detailed microphysics. We present MeerKAT radio continuum and \textsc{H\,i} spectral-line observations of one such normal galaxy, NGC\,1532; a nearby ($D\sim15\,\mathrm{Mpc}$) and edge-on ($i \gtrsim 80^{\circ}$) spiral galaxy tidally interacting with its smaller elliptical companion, NGC\,1531. We find magnetized, highly-ordered radio continuum loops extending $\sim10$\,kpc above and below the disk; visibly connecting discrete star-forming regions in the disk with the nucleus. The deep MeerKAT \textsc{H\,i} observations place an upper limit on the column density of neutral gas coincident with the outflow to $N_\mathrm{\textsc{H\,i}} \lesssim 3 \times 10^{19}\,\mathrm{cm}^{-2}$. Unlike previously observed outflows---for which ejected gas and dust can be traced across multiple wavelengths---the loops in NGC 1532 show no detectable signs of dust or gas coincident with the radio emission far from the disk. We explore multiple possible mechanisms for driving this magnetic wind and favor an explanation where cosmic-ray pressure plays a significant role in launching these outflows.

\end{abstract}



\section{Introduction} \label{sec:intro}

Stellar feedback powers superwinds in galaxies with high star formation rates---expelling metal-rich gas and ionizing radiation into the surrounding circumgalactic medium (CGM). In addition to ejecting gas and metals, galactic winds impart energy and momentum into their surroundings that mix and heat the gas, preventing or delaying the accretion of such gas onto the galaxy to fuel future star-formation episodes. By ejecting material and preventing accretion, winds are critical for regulating galaxy growth and enrichment, and may explain why the stellar masses of $L_*$ galaxies and below fall short of predictions from their primordial gas budget \citep[e.g.,][]{benson03}.

The dominance of different outflow mechanisms is sensitive to the star formation rate surface density $\Sigma_\mathrm{SFR}$ \citep{thompson24}. In archetypal outflow-producing starbursts, $\Sigma_\mathrm{SFR}$ often exceeds $1\,\mathrm{M_\odot\,yr^{-1}\,kpc^{-1}}$. In such intense star formation, the supernovae energize a very hot phase \citep[$T\gtrsim10^7\,\mathrm{K}$;][]{strickland09}, and the super-heated gas expands adiabatically, interacts via ram pressure and/or mixing with surrounding cooler material, and produces a galactic outflow \citep{chevalier85}.  For radiation pressure to drive outflows, a high enough $\Sigma_\mathrm{SFR}$ is needed such that the luminosity exceeds the Eddington luminosity and the ionizing photons can impart their momentum onto surrounding dust grains and drive an outflow. However, in local $L^*$ and below galaxies, the star formation rate surface densities are typically much smaller $\Sigma_\mathrm{SFR}\lesssim0.01\mathrm{M_\odot\,yr^{-1}\,kpc^{-1}}$. 

The consideration of cosmic rays (CRs) to drive outflows has seen a resurgence in the past decade \citep[e.g.][]{hopkins20,chan22,huang22}, in part because they can be more easily coupled to gas---although this is dependent on the transport physics---and have a longer cooling time \citep{thompson24}. In star-forming galaxies, supernovae are the primary accelerators of CR ions, electrons, and positrons. The ions dominate the total CR energy density, and can be constrained most directly through $\gamma$-rays emitted as the CR protons undergo hadronic interactions. However, outside of our Milky Way and a handful of nearby star-forming galaxies, $\gamma$-ray observatories lack the sensitivity to directly constrain the CR energy density in individual galaxies. While CR electrons and positrons contribute little to the total energy density, they are transported by similar mechanisms to their higher energy counterparts. CRe interactions with the galactic magnetic field produce synchrotron radiation. Radio synchrotron thereby provides an easily accessible---albeit less direct---constraint on the CR populations in a wide variety of environments throughout cosmic time.

Simulations now incorporate CRs into stellar feedback models \citep[e.g.,][]{hopkins18} and corroborate earlier theoretical studies that the puffiness of edge-on radio galaxies is dependent on the mode and timescales of CR transport on microphysical scales \citep{boulares90}. Radio observations of edge-on galaxies reveal puffy synchrotron disks with larger vertical scale heights than the non-radio wavelengths \citep[e.g.,][]{krause18}. The Continuum HAlos in Nearby Galaxies -- an EVLA Survey (CHANG-ES) also unveiled surprising characteristics of the magnetic field of radio halos; with some exhibiting reversing magnetic field directions and an X-like morphology seen in the polarized intensity \citep[][]{mora19,stein20,krause20}. The contribution of magnetic field activity to the production of outflows is uncertain, but crucial for utilizing radio synchrotron to disentangle the influence of both CRs and magnetic fields in galaxy growth and evolution.

\cite{parker92} conjectured that CRs are capable of inflating a galactic magnetic field and producing individual magnetic loops that extend out of the galactic plane. Such a signature should be visible via polarization measurements of radio continuum images, but these Parker ``loops'' have been observed in only a handful of galaxies and were coincident with dust emission \citep{beck15}. Recent theoretical studies show that spiral galaxies with strong magnetic field strengths (a few $\mu$G) should be able to drive magnetic outflows via magnetic pressure if there exist processes that amplify the magnetic field on small scales within the disk while maintaining an ordered field on $\sim$kpc scales \citep{steinwandel22}. 

\subsection{NGC 1532: A Normal Star-Forming Galaxy}

NGC\,1532 is a large (diameter $\sim$ 80\,kpc), nearby ($D\sim15$\,Mpc), edge-on ($i\sim83^{\circ}$) spiral galaxy tidally interacting with its smaller elliptical companion, NGC\,1531. 
Apart from its large diameter, NGC\,1532 is a remarkably normal galaxy.
It has a modest star-formation rate (SFR) of $\sim2.7\,\mathrm{M_\odot\,yr^{-1}}$ and stellar mass of $\log\,M_*=10.73\,\mathrm{M_\odot}$, placing it right in the middle of the ``star-forming main-sequence'' \citep{jarrett19}.
The WISE colors \citep[W1-W2 = 0.05 and W3-W4=2.86)][]{jarrett19} and ratio of FIR-radio luminosity \citep[$q_\mathrm{TIR}=2.5$][]{moloko24} suggest star formation processes are dominating the emission coming from NGC\,1532.


We observed NGC\,1532 with MeerKAT while surveying all 298 galaxies in the IRAS Revised Bright Galaxy Sample \citep{sanders03} in the Southern hemisphere \citep{condon21}. 
In a $5\times3$ minute snapshot, the exceptional surface-brightness sensitivity of MeerKAT revealed radio outflows emanating from bright, patchy, star-forming regions.
The magnetized, multi-kpc, CR loops emerging from the disk of NGC~1532 differ from most outflows because they lack detectable signs of stars, dust, or molecular gas \citep{horellou95}. 

The unique and novel radio outflows in conjunction with a lack of detectable dust or cool gas spatially coincident with the synchrotron emission make NGC\,1532 an ideal test subject for studying the effects of CR and magnetic pressure-driven outflows and their contribution to the overall baryon cycle of galaxies. In this work, we present sensitive MeerKAT radio continuum observations of NGC\,1532 and an analysis of the magnetic field strength throughout the disk and radio outflows. Additionally, we introduce deep, high-spectral resolution ($\sim5.6\,\mathrm{km\,s^{-1}}$) \textsc{H\,i} observations of NGC\,1532 taken with the MeerKAT telescope which confirm the absence of cool gas spatially coincident with the synchrotron loops. A detailed analysis of the \textsc{H\,i} kinematics and gas properties of NGC 1532 will be presented in a subsequent work (Matthews et al. 2025, in prep).

\section{Observations}

\begin{figure}
  \centering
  \includegraphics[width=0.47\textwidth]{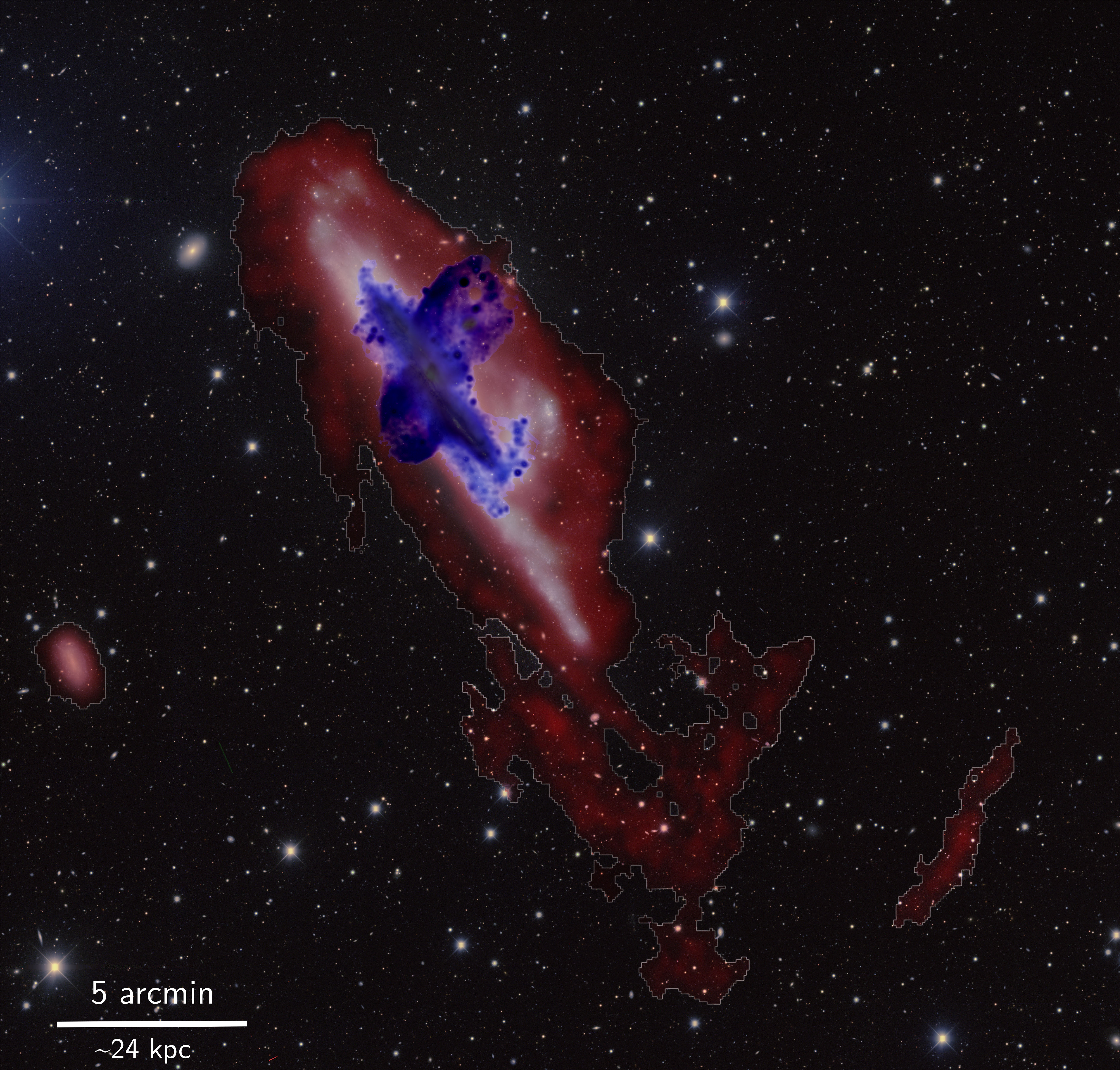}
  \caption{\ion{H}{1} (shaded from dark red to white with increasing intensity) surrounds NGC 1532 and its infalling companion NGC 1531 as well as the orbiting dwarf galaxy ESO~359-G029 (seen on the left border of the image). Extended \ion{H}{1} features---likely the result of tidal interactions amongst the group---are seen far to the southwest of NGC 1532. The 1.3\,GHz radio continuum emission is shown in blue. The background optical image is an RGB image composed of $gri$ filter images from the Dark Energy Survey Data Release 2 \citep{des2}.
  \label{fig:all}}
\end{figure}

The galaxy NGC 1532 was observed using the MeerKAT array in South
Africa \citep{Jonas2016, Camilo2018, DEEP2}
centered at position RA(2000)=4:12:04.31 Dec(2000)=-32:52:59.3 in L
band (856 to 1712 MHz) using at least 58 antennas of the 64 antenna
array. 
Project code DDT-20210802-JC-01 was observed on 7 Aug 2021 for 10 hours,
including calibration and project code SCI-20220822-AM-02 was observed
for 20 hours in 4 sessions between 24 and 31 Oct 2022.
The 2021 observations had 4096 spectral channels across the band and
the 2022 observations had 32768 channels.
The 2022 data were used to obtain HI spectral images and all were used
for the continuum image.
In the 2021 observations, NGC1532 was observed in 10 minute scans
interleaved with calibration observations.  
In the 2022 observations, the on-source scans were 30 minutes in
length. 
All four combinations of the orthogonal linear feeds were recorded.
The photometric calibrator was J0408-6545, the astrometric/gain
calibrator was J0440-4333 and the polarized calibrator J0521+1638
(3C138).

\section{Data Processing}
Processing of the 2021 data followed the general approach of \cite{DEEP2},
\cite{XGalaxy}, and \cite{MK_SMC} and used the Obit
package\footnote{http://www.cv.nrao.edu/$\sim$bcotton/Obit.html}
\citep{Obit}.
The channels needed to image the \ion{H}{1} in NGC 1532 and nearby galaxies
were extracted from the 2022 data, which was also averaged from its native 32K to 4K
channels over the whole band for the continuum image.
The MeerKAT online calibration was applied to the \ion{H}{1} channels and the
channel averaged data were calibrated using the same procedure as the
2021 data.

\subsection{Continuum Calibration}
Calibration and flagging of the continuum data were as described in
\cite{DEEP2} and \cite{MK_SMC}. 
The photometric/bandpass calibrator was J0408-6545, the polarization
calibrator was 3C138 and astrometric calibration used J0440-4333.
The flux density scale was set by the \cite{Reynolds94} spectrum of
PKS~B1934$-$638:
$$
  \log(S) = -30.7667 + 26.4908 \log\bigl(\nu\bigr)
  - 7.0977 \log\bigl(\nu\bigr)^2 $$
  $$+0.605334 \log\bigl(\nu\bigr)^3,$$
where $S$ is the flux density (Jy) and $\nu$ is the frequency (MHz).

\subsection{Continuum Imaging}
           
The combined continuum data from 2021 and 2022 were imaged in Stokes
I, Q, U and V using Obit task MFImage \citep{Cotton2018}.
To accommodate the variation of the sky brightness and the varying
antenna gain with frequency, this imaging used 14 constant fractional
bandwidth (5\%) sub-bands, imaged independently but CLEANed jointly.
The channels with \ion{H}{1} emission were flagged for the Stokes I image.
The curvature of the sky was accommodated using faceting.
Images fully covered the sky to a radius of 1.0$^\circ$ with outlying
facets to 1.5$^\circ$ centered on sources estimated to appear in
excess of 1 mJy from the SUMMS \citep{mau03} or NVSS \citep{NVSS} catalogs.
Two iterations of phase only self calibration were used.
A Briggs Robust factor of -1.5 (AIPS/Obit usage) resulted in a
restoring beam size of 7.5" $\times$ 7.4" at position angle -1$^\circ$.
CLEANing proceeded to a depth of 20 $\mu$Jy/beam in Stokes I, 10
$\mu$Jy/beam in Stokes Q, U and 30 $\mu$Jy/beam in Stokes V.
The off--source RMS in Stokes I is 2.6 $\mu$Jy/beam and 1.4
$\mu$Jy/beam in Stokes Q, U and V.

\subsection{Polarization Processing}

Rotation measure (RM) cubes were derived from the 14-channel deconvolution in Stokes Q and U. The polarized intensity, electric vector position angle (EVPA) at wavelength $\lambda=0$, and RM were determined by a search in Faraday depth in each pixel to find the RM that gave the highest unwrapped polarized intensity. This search was over the range $\pm$150\,rad\,m$^{-2}$ in steps of 0.5 rad\,m$^{-2}$. This process corrects the EVPA for the Faraday rotation.

\subsection{\ion{H}{1} Imaging}\label{sec:h1imaging}
After initial calibration of the continuum datasets as described previously, data from each session were individually corrected for Doppler shifts using Obit task 'CVel', after which they were combined. The initial reduction presented in this paper used only three of the four 2022 datasets. The bandwidth was reduced to 10.5 MHz centered at 1414.8 MHz, keeping $\sim3$\,MHz of line-free continuum on either side of the \ion{H}{1} line. The data were inspected for radio frequency interference (RFI) which was removed, and the \ion{H}{1} line frequencies were masked. After masking, the data was calibrated to the full continuum Stokes I model of the source in phase at 40 second intervals, followed by a calibration in amplitude and phase at 10 minute intervals.  After applying this calibration, the line channels were unmasked and the model was subtracted from the data to remove the continuum emission. To remove any residual continuum emission, the line-free channels immediately above and below the \ion{H}{1} line frequency range were imaged and CLEANed using the Obit task Imager, and the new model was subtracted from the data.  

Each frequency channel in the residual data was imaged independently, again using Obit task Imager, to create a cube of images which span the frequency range from 1412.8 to 1416.8 MHz, or \ion{H}{1} velocities from roughly 1604 km s$^{-1}$ to 762 km s$^{-1}$.  The curvature of the sky was accommodated by using faceting to a radius of 1.0$^\circ$, and the image was CLEANed to a depth of 1.2 $\mathrm{mJy/beam}$. A Briggs robust parameter of 0 (AIPS/OBIT usage) provided a beam of $6.76" \times 6.48"$ at a position angle of $1.86^{\circ}$. The \ion{H}{1} cube was then corrected for the primary beam response.  The final \ion{H}{1} cube has an rms noise of approximately $244\,\mu\mathrm{Jy\,beam^{-1}}$ in each 26.1 kHz channel, corresponding to a velocity resolution of 5.5 km s$^{-1}$.

\section{Global \ion{H}{1} Properties}

The \ion{H}{1} integrated intensity from the MeerKAT pointing centered on NGC\,1532 is shown as a heat map superimposed on an optical image of the region in Figure \ref{fig:all}. We measure the total mass and extent of the \ion{H}{1} gas associated with the NGC\,1532/31 system using the primary-beam corrected \ion{H}{1} cube described in Section \ref{sec:h1imaging}. We used the SoFiA-2 source finder \citep{sofia2} on the \ion{H}{1} cube to create preliminary masks for individual sources. We adjusted the input parameters (e.g. spatial and spectral smoothing-kernel sizes, local signal-to-noise threshold) to SoFiA-2 until the resulting catalog contained the three sources easily identified by-eye: NGC1532/31, ESO 359-G029, and IC 2040. Given the complexity of the emission surrounding NGC 1532, a more robust method of detecting and creating source masks would utilize some combination of automated source-finding algorithms such as SoFiA-2 as well as the virtual-reality datacube viewer iDaViE \citep{marchetti21, jarrett24}. As this work is focused on the radio continuum properties and discovery of unique radio loops, we reserve this more comprehensive treatment of the \ion{H}{1} data for a follow-up work focused on the \ion{H}{1} kinematics (Matthews et al., in prep). 

From our initial analysis, there is $1.8\pm0.2\times10^{10}\,\mathrm{M_\odot}$ of \ion{H}{1} associated with NGC1532---excluding the diffuse emission extending to the southwest. The \ion{H}{1} stretches $\sim18\arcmin$ ($\sim$87\,kpc) along NGC 1532's major axis and $\sim6\arcmin$ ($\sim29$\,kpc) along the minor axis.
We smoothed the \ion{H}{1} data to a beam size of $20''\times20''$ to increase the sensitivity around the radio loops. Within a region 8\,kpc above the disk midplane and coincident with the radio continuum loops, we constrain the column density to $N_\mathrm{HI}\lesssim 3\times 10^{19}\,\mathrm{cm^{-2}}$. Notably, this is two orders of magnitude less than the column density of \ion{H}{1} associated with the outflows in M82 \citep{martini18}.



\section{Radio Continuum Properties}




\subsection{Spectral Index}
\begin{figure}[hb]
  \centering
  \includegraphics[trim={0.5cm 0cm 1cm 0cm}, clip, width=0.49\textwidth]{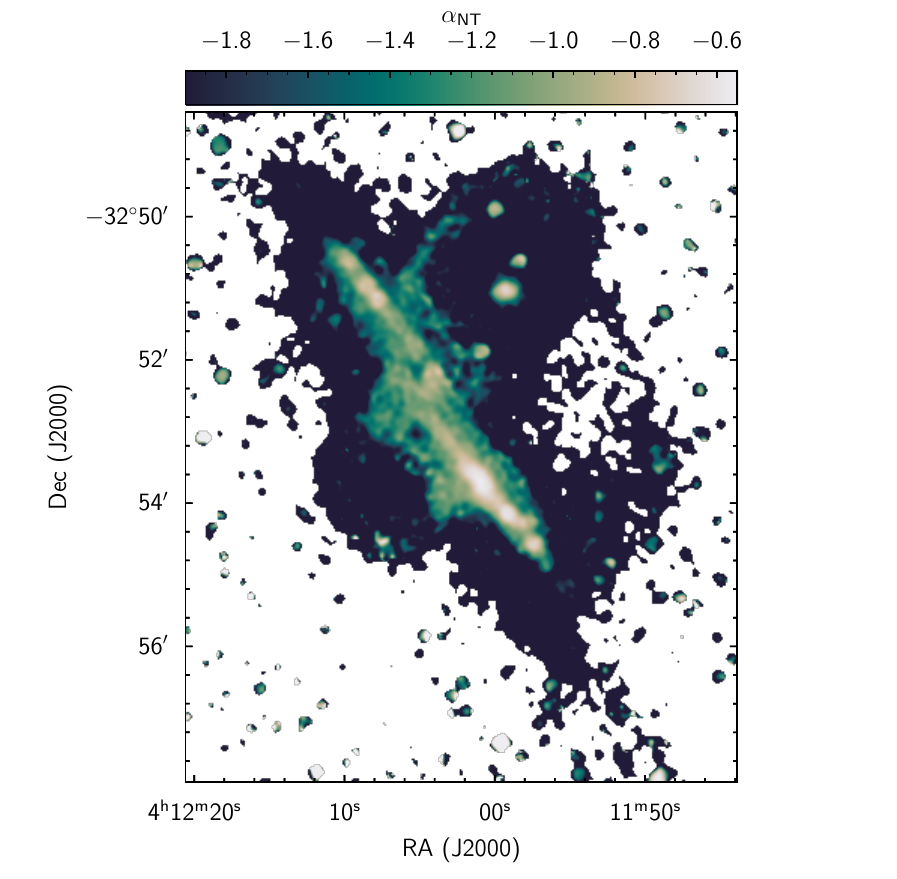}
  \caption{The non-thermal spectral index $\alpha_\mathrm{NT}$ of NGC 1532 is shown in colorscale.
    \label{fig:alpha}}
\end{figure}

We calculate the observed spectral index $\alpha_\mathrm{total}$ (adopting the sign convention $S\propto\nu^{\alpha}$) for each pixel as the slope of the best-fitting line to the relationship between $\log\,S$ and $\log\,\nu$ across the 14 frequency subbands. The non-thermal spectral index $\alpha_\mathrm{NT}$ can be estimated from
\begin{equation}
\alpha_\mathrm{NT} = \alpha_\mathrm{total}\left(1+\frac{S_\mathrm{T}}{S}\right) + \alpha_\mathrm{T}\left(\frac{S_\mathrm{T}}{S}\right),
\end{equation}
where $\alpha_\mathrm{T} = -0.1$ is the spectral index of the thermal radio emission and $S_\mathrm{T}/S = 0.08$ is the thermal fraction at $\sim$1\,GHz \citep{niklas97}. The resulting non-thermal spectral indices are shown in Figure \ref{fig:alpha} for the disk of NGC\,1532 and its companion NGC\,1531. Since the lack of zero-spacings more strongly limits the surface-brightness sensitivity of the higher frequency channels, the spectral index for diffuse emission appears artificially steeper than the true physical value. For this reason, while we display the calculated spectral indices over all of NGC 1532 in Figure \ref{fig:alpha}, we restrict our use of the direct measurement of $\alpha_\mathrm{NT}$ to the disk and assume a typical value of $\alpha_\mathrm{NT}=-1$ for the diffuse emission related to the loops.

In the intensely star-forming complex in the south-western part of the disk, $\alpha_\mathrm{NT}\sim-0.55$. This is likely a region with increased thermal contribution to the $1.3\,$GHz emission and will be characterized with planned H$\alpha$ observations. This shallow spectral index is also consistent with very young electrons recently accelerated in the shocks of supernova remnants.

The nuclear and north-eastern star-forming regions have average $\alpha_\mathrm{NT} \sim -0.8$, close to the value $\langle\alpha_\mathrm{NT}\rangle=-0.83$ found for a typical sample of spiral galaxies in the local universe \citep{niklas97}. The lack of a compact, flat-spectrum ($\alpha\sim-0.5$) source in the nucleus of NGC\,1532 suggests there is not currently a young, active AGN in its core \citep[e.g.][]{degasperin18}, but this does not preclude the possibility of past AGN activity.

\subsection{Polarization Properties and Morphology}

\begin{figure}
  \centering
  \includegraphics[trim={2.5cm 3cm 1.9cm 1.6cm}, clip, width=0.47\textwidth]{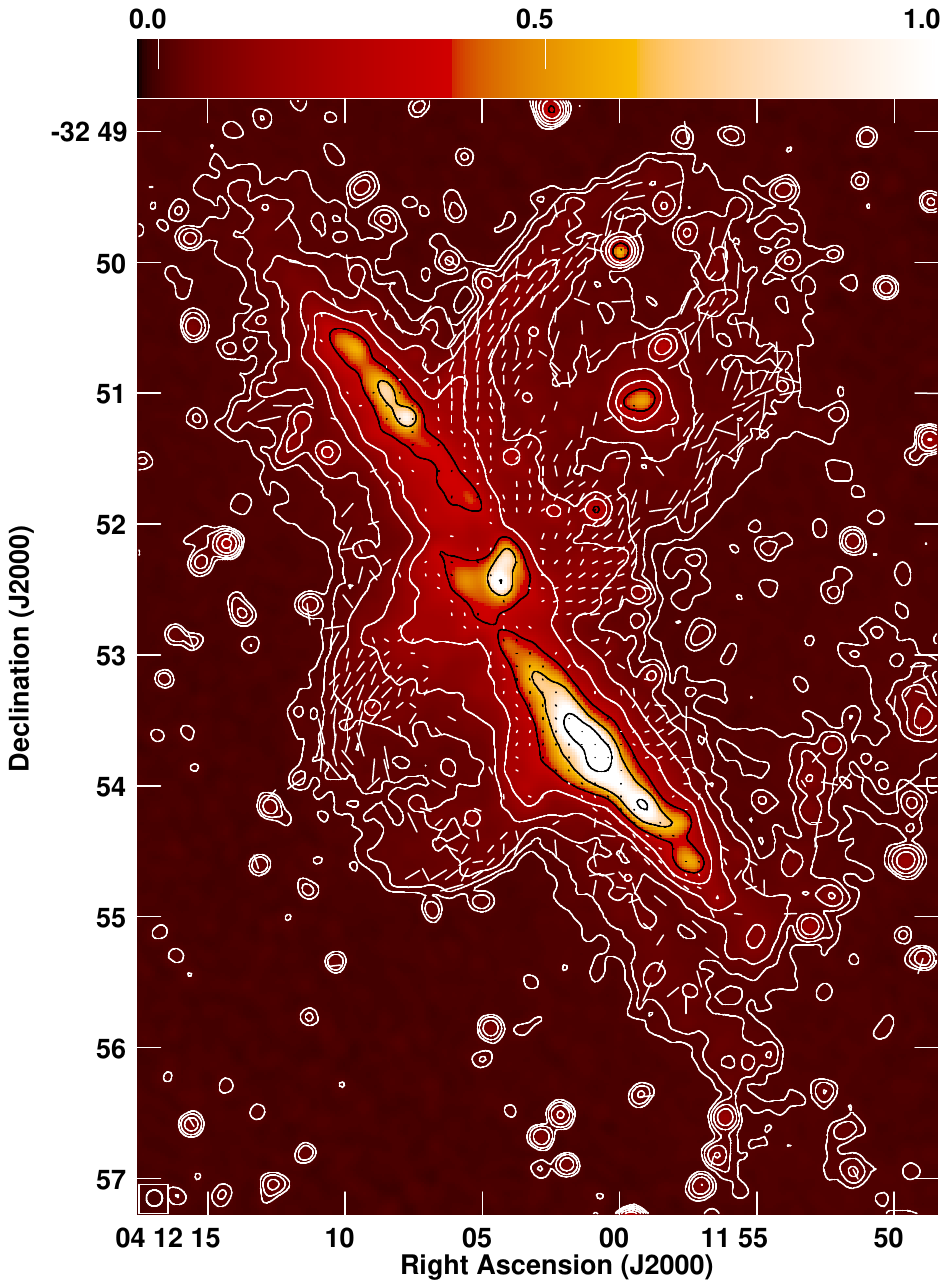}
  \caption{Polarization vectors (rotated by $90^\circ$ to correspond to the direction of the magnetic field) are superimposed on the grayscale radio continuum background of NGC 1532. White contour levels of the continuum flux density are $10\,\mu\mathrm{Jy\,beam^{-1}}\times(1,2,4,8,16,32,64,128)$. The length of the polarization vectors correspond to the polarization intensity such that a 20$''$ vector corresponds to a polarization intensity of 0.08 Jy$\,\mathrm{beam}^{-1}$. 
  }\label{fig:pol}
\end{figure}

The linearly polarized emission---where the vectors have been rotated by 90$^{\circ}$ to reflect the direction of the magnetic field---of NGC 1532 is shown in Figure \ref{fig:pol}. The magnetic field is oriented vertically relative to the disk midplane. At the maximum height of their visible extent, the magnetic field traces the edge of the radio synchrotron plumes, appearing to reconnect $\sim$8--12\,kpc above/below the disk.

\begin{figure}
  \centering
  \includegraphics[trim={0.4cm 1.5cm 0cm 1cm},clip,width=0.49\textwidth]{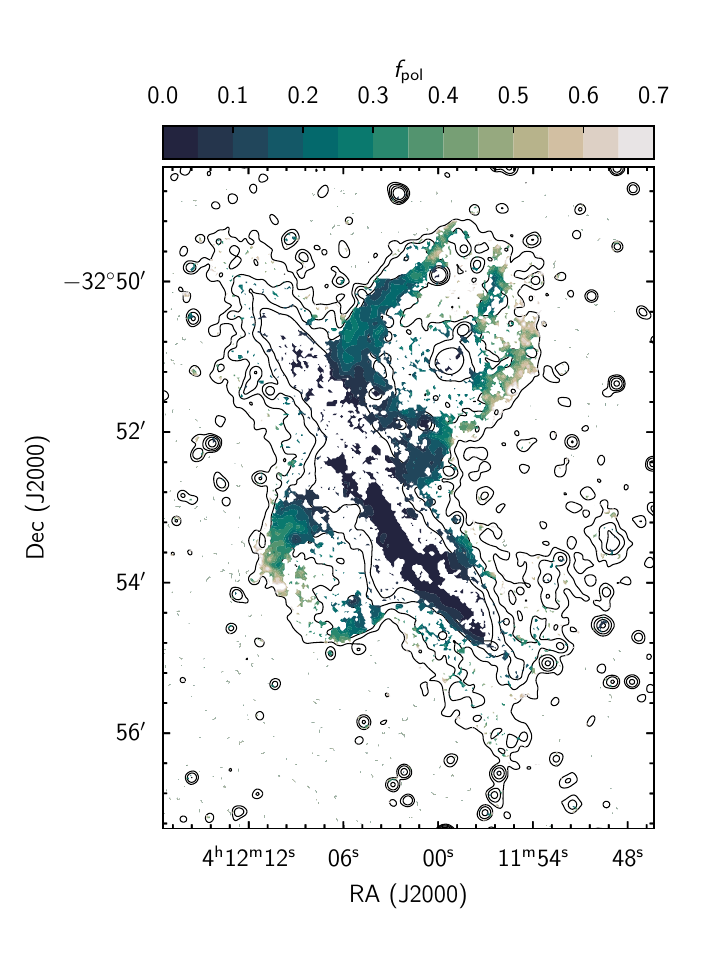}
  \caption{The fraction of polarized radio continuum to total intensity (restricted to regions with $>1\sigma$ in polarized intensity) shown in colorscale. The degree of polarization reflects the degree to which the magnetic field is ordered at that location. The radio continuum is shown as the black contours with levels at $10\,\mu\mathrm{Jy\,beam^{-1}}\times\,$(1, 2, 4, 8, 16). 
    \label{fig:fpol}}
\end{figure}

The fraction of the total synchrotron emission that is polarized reflects the degree of order within the magnetic field. As seen in Figure \ref{fig:fpol}, the field is less ordered within the disk ($<10\%$), characteristic of small-scale turbulence induced by ongoing star formation activity. The degree of polarization increases with distance from the disk, likely due to a combination of a decreasing thermal contribution to the 1.3\,GHz flux as well as a tendency for magnetic fields to be more ordered on larger scales \citep{krause20}.


\subsection{Magnetic Field Strength}

\begin{figure}
  \centering
  \includegraphics[trim={1mm 1mm 1cm 1cm},clip,width=0.49\textwidth]{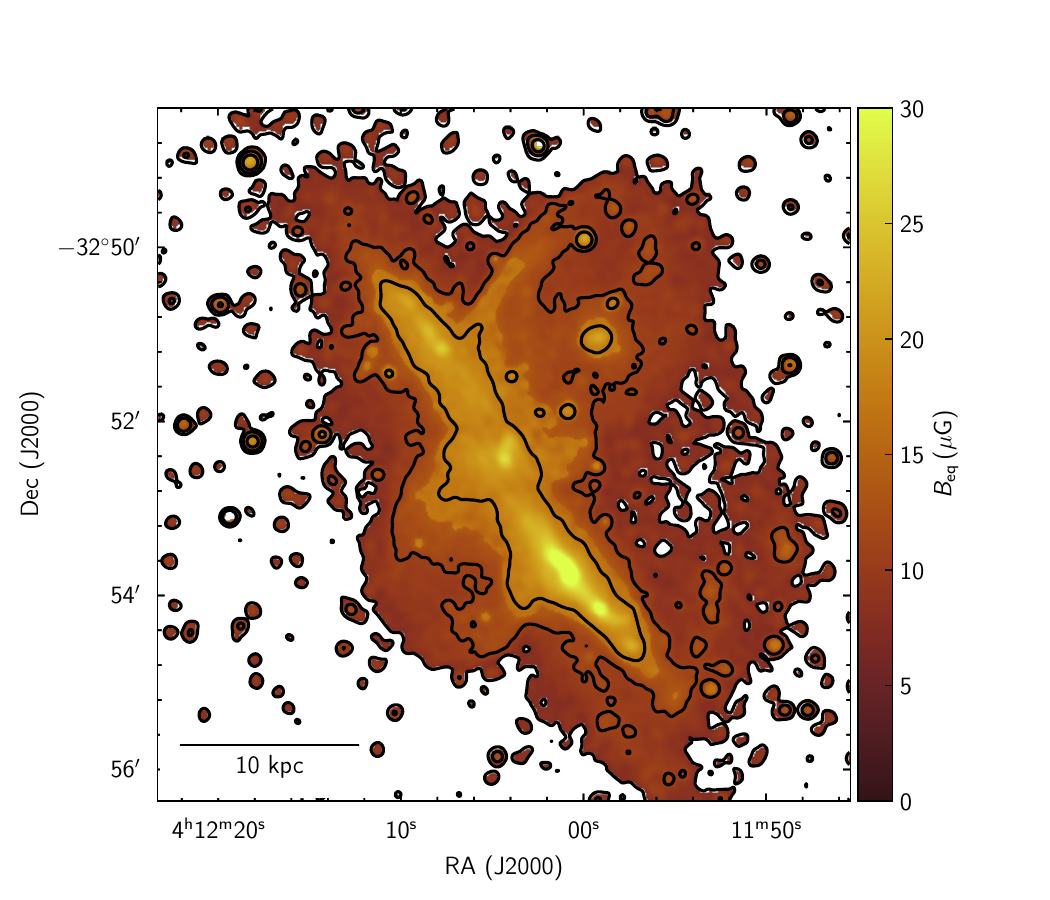}
  \caption{Equipartition magnetic field strength calculated using the revised formula from \cite{beck05} spans the range $20-35\,\mu$G in the disk and $10-15\,\mu$G in the halo.
    \label{fig:bfield}}
\end{figure}

We assume the CR and magnetic energy densities are in equipartition, which although imperfect, enables us to estimate the magnetic field strength in the disk and outflow of NGC\,1532. We use the revised formula from \cite{beck05},
\begin{eqnarray}
\nonumber  B_\mathrm{eq} = \{ 4\pi(2\alpha+1)(\mathbf{K_0} +1) I_\nu E_\mathrm{p}^{1-2\alpha} (\nu/2c_1)^{\alpha} \\
  / \ [(2\alpha-1)c_2(\alpha)l\,c_4(i)\}^{1/(\alpha+3)},
\end{eqnarray}
where $\alpha$ is the synchrotron spectral index, $I_\nu$ is the synchrotron intensity at frequency $\nu$, $\mathbf{K_0}$ is the constant ratio of the proton to electron number densities, $E_\mathrm{p}$ is the proton rest energy, and $l$ is the pathlength. The constants $c_1, c_2,$ and $c_4$ depend on $\alpha$ and the inclination $i$. We refer the reader to the Appendix of \cite{beck05} for a detailed derivation. We use the measured spectral index $\alpha_\mathrm{NT}$ when calculating the magnetic field for the smaller-spatial scale emission in the disk and its immediate vicinity, but adopt a typical value of $\alpha_\mathrm{NT}\sim-1$ for the diffuse loop. For the pathlength $l$ of the electrons, we assume 1\,kpc/cos\,$i$, where $i\sim83^\circ$.

The equipartition magnetic field strength in the disk ranges from 20$\,\mu$G in the quiescent regions to $35\,\mu$G in the star-forming complex in the southwestern region of the disk (see Figure \ref{fig:bfield}). The magnetic field decreases within the radio loops to a range between 10$\,\mu$G and 15$\,\mu$G. Assuming equipartition likely overestimates the magnetic field in the radio loops, however, the CHANG-ES survey \citep{irwin24} using the same method finds the average magnetic field in the halos of edge-on galaxies to be $B_\mathrm{halo}\sim5\,\mu$G, 30--50\% of the value observed in NGC\,1532.

\section{The nature of the radio loops}

The unique radio features of NGC\,1532 expose some of the most uncertain physics of galactic superwinds: how capable is CR (or magnetic) pressure to influence the occurence and nature of outflows, how much material---and in what phase (e.g. hot, warm-ionized, or cold neutral)---can CR-driven outflows expel, how do superwinds impact the CGM? The star formation rate surface density $\Sigma_\mathrm{SFR}$ of NGC 1532 is orders of magnitude less than that of galaxies driving biconical outflows from the nucleus (e.g., M82), yet exhibits prominent, magnetized CR loops many kpc off the disk midplane. In this section we explore possible mechanisms for generating the magnetized outflows and outline future questions we will address in a sequence of following papers.

\subsection{Primary accelerator of cosmic-ray electrons}

The base of the radio loops appear connected with discrete star-forming regions in both the disk and nucleus. Closer examination of the southwestern star-forming complex in particular reveals polarization vectors directly emanating from the disk into the halo of NGC 1532 (see Figure \ref{fig:pol}). The connection is strongest for the polarized vectors directed north of this region, but is seen in the south as well, where the weaker connection may be explained by a projection effect due to the inclination of NGC 1532 being less than 90$^\circ$.

AGN are well-known drivers of synchrotron emission from the cores of galaxies. The ratio of infrared to radio emission for NGC 1532 as well as its location on a WISE color-color plot is consistent with star-forming galaxies \cite{jarrett19,galante24, moloko24}. Our measurement of the non-thermal spectral index does not reveal a compact core with a flattened spectral index, which further indicates a lack of ongoing AGN activity. While there is no indication of an AGN in NGC 1532, we cannot rule out the presence of AGN activity in the past.

\subsection{Physical mechanism of outflow}

Regardless of the CR source being star formation or an AGN, the physical mechanism driving the outflow could be one or a combination of the following: thermal, radiation, magnetic, and cosmic-ray pressure. Thermal pressure is the mechanism thought to be primarily responsible for the galactic superwinds of M82; the theory behind which was originally presented by \cite{chevalier85}. In a dense starburst like M82, supernovae energize a very hot phase, where---given a characteristic temperature, density, and velocity---the super-heated gas expands adiabatically and interacts via ram pressure and/or mixing with surrounding cooler material. Hard X-ray emission, particularly line emission, is a strong observational signpost of this thermal wind model, and implies gas temperatures in M82's core of $T\gtrsim 5\times 10^{7}\mathrm{K}$ \citep{strickland09}.

Unlike M82, NGC 1532 has a modest SFR $\sim2.7\,\mathrm{M_\odot\,yr^{-1}}$ derived from its $12\,\mu$m WISE luminosity \cite{jarrett19}. 
Unlike other galactic outflows emanating from galaxies of similar SFRs (e.g. NGC 253), the majority of the star formation is occurring in the disk of the galaxy instead of the nucleus. 
In the most active star-forming region located southwest of the nucleus, the star formation rate surface density $\Sigma_\mathrm{SFR}$ is at most $\sim0.1\,\mathrm{M_\odot\,yr^{-1}\,kpc^{-2}}$, calculated from the radio continuum luminosity. 
Assuming Milky Way values for the dust properties, \cite{thompson24} estimate $\Sigma_\mathrm{SFR}\gtrsim0.2\,\mathrm{M_\odot\,yr^{-1}\,kpc^{-2}}$ for the luminosity from star-formation to exceed the Eddington luminosity. Exceeding this threshold is required for radiation pressure to exert enough momentum on the dust grains to drive outflows.
It therefore appears unlikely that radiation pressure alone (on dust, specifically) is driving the observed outflows.

A combination of several physical mechanisms likely contribute to driving galactic superwinds rather than a singular driver. \cite{buckman20} investigated the role of CRs on driving outflows in M82 and found that strong hadronic losses limit the CR energy density in M82's core such that it is dynamically weak with respect to gravity, and largely incapable of driving outflows. NGC\,1532 has smaller gas and star-formation rate surface densities where CR protons have weaker pion cooling and are able to achieve dynamically important energy densities and pressures to drive galactic winds. A careful treatment of thermal, radiative, and magnetic pressures is needed to quantify the role of CRs in NGC 1532's outflow and will be presented in a subsequent paper following the completion of necessary observations. 




\section{Summary}

We present new MeerKAT continuum imaging at 1.3\,GHz and high-resolution \ion{H}{1} spectral imaging of galaxy NGC~1532---revealing a galactic-scale, magnetized outflow of cosmic-ray electrons. We summarize the first results from these observations below.

\begin{itemize}
\item The loops of cosmic-ray electrons appear to originate from discrete star-forming regions in the disk and the nucleus and extend $\sim$10\,kpc out of disk mid-plane. There is no detectable dust along the radio loops, and we place a strong upper limit on the neutral hydrogen column density $\sim8\,$kpc above the disk, coincident with the synchrotron loops, of $N_\mathrm{HI}<3\times10^{19}\,\mathrm{cm^{-1}}$. 
\item The calculated magnetic field strength, assuming the revised equipartition formulas of \cite{beck05}, in the disk and the halo is stronger than that of other edge-on star-forming galaxies \citep{krause20}. Within the most actively star-forming complex---and ergo the most turbulent---the total magnetic field strength reaches $35\,\mu$G. In the radio loops, the magnetic field strength is 10-15$\,\mu$G---even 10\,kpc off the disk midplane.

\item NGC 1532 is an exquisite laboratory to learn about the role of cosmic-rays in galactic feedback. NGC~1532 has a moderate $\Sigma_\mathrm{SFR}\sim0.01-0.1\,\mathrm{M_\odot\,yr^{-1}\,kpc^{-2}}$---in the regime of activity where CR pressure is theorized to play a significant role in driving multi-kpc outflows. The proximity to us makes it possible to trace the radio outflows to discrete regions within the disk where the local SFR and gas conditions differ from the disk average.
\item Further observations of the (potential) warm-ionized and hot phases of the outflow are necessary to calculate and compare the thermal, radiative, CR, and magnetic pressures within the outflow. Doing so will determine whether the winds of the normal, non-starbursting NGC 1532 are the first conclusive example of primarily CR-driven winds.
  \item Examples of ordered features in radio halos are accumulating \citep[e.g.,][]{heesen24}. The exquisite surface-brightness sensitivity of MeerKAT and future radio interferometers will enable detailed studies of CRs in galactic superwinds over a large range of galaxy types and environments.
\end{itemize}


\begin{acknowledgments}
AMM thanks Dr. Gwen Rudie for countless discussions and lending her comprehensive knowledge of stellar feedback and the gas in and around galaxies---this work is far richer because of it. The MeerKAT telescope is operated by the South African Radio Astronomy Observatory, which is a facility of the National Research Foundation, an agency of the Department of Science and Innovation. The National Radio Astronomy Observatory is a facility of the National Science Foundation, operated under a cooperative agreement by Associated Universities, Inc. WMP acknowledges that basic research in radio astronomy at the U.S. Naval Research Laboratory is supported by 6.1 Base funding. LM acknowledges financial support from the Inter-University Institute for Data Intensive Astronomy (IDIA), a partnership of the University of Cape Town, the University of Pretoria and the University of the Western Cape, and from the South African Department of Science and Innovation’s National Research Foundation under the ISARP RADIOMAP Joint Research Scheme (DSI-NRF Grant Number 150551) and the CPRR Projects (DSI-NRF Grant Number SRUG2204254729).
\end{acknowledgments}

%

\vspace{5mm}
\facilities{MeerKAT}






\bibliography{refs}

\begin{thebibliography}{}
\expandafter\ifx\csname natexlab\endcsname\relax\def\natexlab#1{#1}\fi

\bibitem[{{Abbott} {et~al.}(2021){Abbott}, {Adam{\'o}w}, {Aguena}, {Allam},
  {Amon}, {Annis}, {Avila}, {Bacon}, {Banerji}, {Bechtol}, {Becker},
  {Bernstein}, {Bertin}, {Bhargava}, {Bridle}, {Brooks}, {Burke}, {Carnero
  Rosell}, {Carrasco Kind}, {Carretero}, {Castander}, {Cawthon}, {Chang},
  {Choi}, {Conselice}, {Costanzi}, {Crocce}, {da Costa}, {Davis}, {De Vicente},
  {DeRose}, {Desai}, {Diehl}, {Dietrich}, {Drlica-Wagner}, {Eckert},
  {Elvin-Poole}, {Everett}, {Evrard}, {Ferrero}, {Fert{\'e}}, {Flaugher},
  {Fosalba}, {Friedel}, {Frieman}, {Garc{\'\i}a-Bellido}, {Gaztanaga},
  {Gelman}, {Gerdes}, {Giannantonio}, {Gill}, {Gruen}, {Gruendl}, {Gschwend},
  {Gutierrez}, {Hartley}, {Hinton}, {Hollowood}, {Honscheid}, {Huterer},
  {James}, {Jeltema}, {Johnson}, {Kent}, {Kron}, {Kuehn}, {Kuropatkin},
  {Lahav}, {Li}, {Lidman}, {Lin}, {MacCrann}, {Maia}, {Manning}, {Maloney},
  {March}, {Marshall}, {Martini}, {Melchior}, {Menanteau}, {Miquel}, {Morgan},
  {Myles}, {Neilsen}, {Ogando}, {Palmese}, {Paz-Chinch{\'o}n}, {Petravick},
  {Pieres}, {Plazas}, {Pond}, {Rodriguez-Monroy}, {Romer}, {Roodman}, {Rykoff},
  {Sako}, {Sanchez}, {Santiago}, {Scarpine}, {Serrano}, {Sevilla-Noarbe},
  {Smith}, {Smith}, {Soares-Santos}, {Suchyta}, {Swanson}, {Tarle}, {Thomas},
  {To}, {Tremblay}, {Troxel}, {Tucker}, {Turner}, {Varga}, {Walker},
  {Wechsler}, {Weller}, {Wester}, {Wilkinson}, {Yanny}, {Zhang}, {Nikutta},
  {Fitzpatrick}, {Jacques}, {Scott}, {Olsen}, {Huang}, {Herrera}, {Juneau},
  {Nidever}, {Weaver}, {Adean}, {Correia}, {de Freitas}, {Freitas},
  {Singulani}, {Vila-Verde}, \& {Linea Science Server}}]{des2}
{Abbott}, T.~M.~C., {Adam{\'o}w}, M., {Aguena}, M., {et~al.} 2021, \apjs, 255,
  20

\bibitem[{{Beck}(2015)}]{beck15}
{Beck}, R. 2015, A\&A, 578, A93

\bibitem[{{Beck} \& {Krause}(2005)}]{beck05}
{Beck}, R., \& {Krause}, M. 2005, Astronomische Nachrichten, 326, 414

\bibitem[{{Benson} {et~al.}(2003){Benson}, {Bower}, {Frenk}, {Lacey}, {Baugh},
  \& {Cole}}]{benson03}
{Benson}, A.~J., {Bower}, R.~G., {Frenk}, C.~S., {et~al.} 2003, \apj, 599, 38

\bibitem[{{Boulares} \& {Cox}(1990)}]{boulares90}
{Boulares}, A., \& {Cox}, D.~P. 1990, \apj, 365, 544

\bibitem[{{Buckman} {et~al.}(2020){Buckman}, {Linden}, \&
  {Thompson}}]{buckman20}
{Buckman}, B.~J., {Linden}, T., \& {Thompson}, T.~A. 2020, \mnras, 494, 2679

\bibitem[{Camilo {et~al.}(2018)Camilo, Scholz, Serylak, \& et~al.}]{Camilo2018}
Camilo, F., Scholz, P., Serylak, M., \& et~al. 2018, ApJ, 856, 180

\bibitem[{{Chan} {et~al.}(2022){Chan}, {Kere{\v{s}}}, {Gurvich}, {Hopkins},
  {Trapp}, {Ji}, \& {Faucher-Gigu{\`e}re}}]{chan22}
{Chan}, T.~K., {Kere{\v{s}}}, D., {Gurvich}, A.~B., {et~al.} 2022, \mnras, 517,
  597

\bibitem[{{Chevalier} \& {Clegg}(1985)}]{chevalier85}
{Chevalier}, R.~A., \& {Clegg}, A.~W. 1985, Nat, 317, 44

\bibitem[{Condon {et~al.}(1998)Condon, Cotton, Greisen, Yin, Perley, Taylor, \&
  Broderick}]{NVSS}
Condon, J.~J., Cotton, W.~D., Greisen, E.~W., {et~al.} 1998, AJ, 115, 1693

\bibitem[{{Condon} {et~al.}(2021){Condon}, {Cotton}, {Jarrett},
  {et~al.}}]{condon21}
{Condon}, J.~J., {Cotton}, W.~D., {Jarrett}, T., {et~al.} 2021, ApJS, 257, 35

\bibitem[{{Cotton}(2008)}]{Obit}
{Cotton}, W.~D. 2008, \pasp, 120, 439

\bibitem[{Cotton {et~al.}(2018)Cotton, Condon, Kellermann, Lacy, Perley,
  Matthews, Vernstrom, Scott, \& Wall}]{Cotton2018}
Cotton, W.~D., Condon, J.~J., Kellermann, K.~I., {et~al.} 2018, ApJ, 856, 67

\bibitem[{Cotton {et~al.}(2020)Cotton, Thorat, Condon, Frank, Józsa, White,
  Deane, Oozeer, Atemkeng, Bester, Fanaroff, Kupa, Smirnov, Mauch, Krishnan, \&
  Camilo}]{XGalaxy}
Cotton, W.~D., Thorat, K., Condon, J.~J., {et~al.} 2020, MNRAS, 495, 1271

\bibitem[{{Cotton} {et~al.}(2024){Cotton}, {Filipovi{\'c}}, {Camilo},
  {Indebetouw}, {Alsaberi}, {Anih}, {Baker}, {Bastian }, {Boji{\v{c}}i{\'c}},
  {Carli}, {Cavallaro}, {Crawford}, {Dai}, {Haberl}, {Levin}, {Luken}, {Pennoc
  k}, {Rajabpour}, {Stappers}, {van Loon}, {Zijlstra}, {Buchner}, {Geyer},
  {Goedhart}, \& {Serylak}}]{MK_SMC}
{Cotton}, W.~D., {Filipovi{\'c}}, M.~D., {Camilo}, F., {et~al.} 2024, \mnras,
  529, 2443

\bibitem[{{de Gasperin} {et~al.}(2018){de Gasperin}, {Intema}, \&
  {Frail}}]{degasperin18}
{de Gasperin}, F., {Intema}, H.~T., \& {Frail}, D.~A. 2018, \mnras, 474, 5008

\bibitem[{{Galante} {et~al.}(2024){Galante}, {Saponara}, {Romero}, \&
  {Benaglia}}]{galante24}
{Galante}, C.~A., {Saponara}, J., {Romero}, G.~E., \& {Benaglia}, P. 2024,
  \aap, 685, A157

\bibitem[{{Heesen} {et~al.}(2024){Heesen}, {Wiegert}, {Irwin}, {Crocker},
  {Kiehn}, {Li}, {Wang}, {Stein}, {Dettmar}, {Soida}, {Henriksen}, {Gajovic},
  {Yang}, \& {Br{\"u}ggen}}]{heesen24}
{Heesen}, V., {Wiegert}, T., {Irwin}, J., {et~al.} 2024, arXiv e-prints,
  arXiv:2409.15449

\bibitem[{{Hopkins} {et~al.}(2018){Hopkins}, {Wetzel}, {Kere{\v{s}}},
  {et~al.}}]{hopkins18}
{Hopkins}, P.~F., {Wetzel}, A., {Kere{\v{s}}}, D., {et~al.} 2018, MNRAS, 480,
  800

\bibitem[{{Hopkins} {et~al.}(2020){Hopkins}, {Chan}, {Garrison-Kimmel}, {Ji},
  {Su}, {Hummels}, {Kere{\v{s}}}, {Quataert}, \&
  {Faucher-Gigu{\`e}re}}]{hopkins20}
{Hopkins}, P.~F., {Chan}, T.~K., {Garrison-Kimmel}, S., {et~al.} 2020, \mnras,
  492, 3465

\bibitem[{{Horellou} {et~al.}(1995){Horellou}, {Casoli}, \&
  {Dupraz}}]{horellou95}
{Horellou}, C., {Casoli}, F., \& {Dupraz}, C. 1995, A\&A, 303, 361

\bibitem[{{Huang} \& {Davis}(2022)}]{huang22}
{Huang}, X., \& {Davis}, S.~W. 2022, \mnras, 511, 5125

\bibitem[{{Irwin} {et~al.}(2024){Irwin}, {Beck}, {Cook}, {Dettmar}, {English},
  {Heesen}, {Henriksen}, {Jiang}, {Li}, {Lu}, {Mele}, {M{\"u}ller}, {Murphy},
  {Porter}, {Rand}, {Skeggs}, {Stein}, {Stein}, {Stil}, {Strong}, {Walterbos},
  {Wang}, {Wiegert}, \& {Yang}}]{irwin24}
{Irwin}, J., {Beck}, R., {Cook}, T., {et~al.} 2024, Galaxies, 12, 22

\bibitem[{{Jarrett} {et~al.}(2024){Jarrett}, {Comrie}, {Sivitilli},
  {Pretorius}, {Vitello}, \& {Marchetti}}]{jarrett24}
{Jarrett}, T., {Comrie}, A., {Sivitilli}, A., {et~al.} 2024, in Zenodo
  Software, Vol.~46 (Zenodo), 4614115

\bibitem[{{Jarrett} {et~al.}(2019){Jarrett}, {Cluver}, {Brown},
  {et~al.}}]{jarrett19}
{Jarrett}, T.~H., {Cluver}, M.~E., {Brown}, M.~J.~I., {et~al.} 2019, ApJS, 245,
  25

\bibitem[{Jonas(2016)}]{Jonas2016}
Jonas, J. 2016, ``'' (SARAO), 1, doi:10.22323/1.277.0001

\bibitem[{{Krause} {et~al.}(2018){Krause}, {Irwin}, {Wiegert}, {Miskolczi},
  {Damas-Segovia}, {Beck}, {Li}, {Heald}, {M{\"u}ller}, {Stein}, {Rand},
  {Heesen}, {Walterbos}, {Dettmar}, {Vargas}, {English}, \&
  {Murphy}}]{krause18}
{Krause}, M., {Irwin}, J., {Wiegert}, T., {et~al.} 2018, \aap, 611, A72

\bibitem[{{Krause} {et~al.}(2020){Krause}, {Irwin}, {Schmidt}, {Stein},
  {Miskolczi}, {Carolina Mora-Partiarroyo}, {Wiegert}, {Beck}, {Stil}, {Heald},
  {Li}, {Damas-Segovia}, {Vargas}, {Rand}, {West}, {Walterbos}, {Dettmar},
  {English}, \& {Woodfinden}}]{krause20}
{Krause}, M., {Irwin}, J., {Schmidt}, P., {et~al.} 2020, \aap, 639, A112

\bibitem[{{Marchetti} {et~al.}(2022){Marchetti}, {Jarrett}, {Comrie},
  {Sivitilli}, {Vitello}, {Becciani}, \& {Taylor}}]{marchetti21}
{Marchetti}, L., {Jarrett}, T., {Comrie}, A., {et~al.} 2022, in Astronomical
  Society of the Pacific Conference Series, Vol. 532, Astronomical Data
  Analysis Software and Systems XXX, ed. J.~E. {Ruiz}, F.~{Pierfedereci}, \&
  P.~{Teuben}, 85

\bibitem[{{Martini} {et~al.}(2018){Martini}, {Leroy}, {Mangum}, {Bolatto},
  {Keating}, {Sandstrom}, \& {Walter}}]{martini18}
{Martini}, P., {Leroy}, A.~K., {Mangum}, J.~G., {et~al.} 2018, \apj, 856, 61

\bibitem[{{Mauch} {et~al.}(2003){Mauch}, {Murphy}, {Buttery}, {Curran},
  {Hunstead}, {Piestrzynski}, {Robertson}, \& {Sadler}}]{mau03}
{Mauch}, T., {Murphy}, T., {Buttery}, H.~J., {et~al.} 2003, MNRAS, 342, 1117

\bibitem[{{Mauch} {et~al.}(2020){Mauch}, {Cotton}, {Condon}, {Matthews},
  {Abbott}, {Adam}, {Aldera}, {Asad}, {Bauermeister}, {Bennett}, {Bester},
  {Botha}, {Brederode}, {Brits}, {Buchner}, {Burger}, {Camilo}, {Chalmers},
  {Cheetham}, {de Villiers}, {de Villiers}, {Dikgale-Mahlakoana}, {du Toit},
  {Esterhuyse}, {Fadana}, {Fanaroff}, {Fataar}, {February}, {Frank},
  {Gamatham}, {Geyer}, {Goedhart}, {Gounden}, {Gumede}, {Heywood}, {Hlakola},
  {Horrell}, {Hugo}, {Isaacson}, {J{\'o}zsa}, {Jonas}, {Julie}, {Kapp},
  {Kasper}, {Kenyon}, {Kotz{\'e}}, {Kriek}, {Kriel}, {Kusel}, {Lehmensiek},
  {Loots}, {Lord}, {Lunsky}, {Madisa}, {Magnus}, {Main}, {Malan}, {Manley},
  {Marais}, {Martens}, {Merry}, {Millenaar}, {Mnyandu}, {Moeng}, {Mokone},
  {Monama}, {Mphego}, {New}, {Ngcebetsha}, {Ngoasheng}, {Ockards}, {Oozeer},
  {Otto}, {Patel}, {Peens-Hough}, {Perkins}, {Ramaila}, {Ramudzuli}, {Renil},
  {Richter}, {Robyntjies}, {Salie}, {Schollar}, {Schwardt}, {Serylak},
  {Siebrits}, {Sirothia}, {Smirnov}, {Sofeya}, {Stone}, {Taljaard}, {Tasse},
  {Theron}, {Tiplady}, {Toruvanda}, {Twum}, {van Balla}, {van der Byl}, {van
  der Merwe}, {Van Tonder}, {Wallace}, {Welz}, {Williams}, \& {Xaia}}]{DEEP2}
{Mauch}, T., {Cotton}, W.~D., {Condon}, J.~J., {et~al.} 2020, ApJ, 888, 61

\bibitem[{{Moloko} {et~al.}(2024){Moloko}, {Marchetti}, {Jarrett},
  {et~al.}}]{moloko24}
{Moloko}, M., {Marchetti}, L., {Jarrett}, T., {et~al.} 2024, submitted to MNRAS

\bibitem[{{Mora-Partiarroyo} {et~al.}(2019){Mora-Partiarroyo}, {Krause},
  {Basu}, {Beck}, {Wiegert}, {Irwin}, {Henriksen}, {Stein}, {Vargas}, {Heesen},
  {Walterbos}, {Rand}, {Heald}, {Li}, {Kamieneski}, \& {English}}]{mora19}
{Mora-Partiarroyo}, S.~C., {Krause}, M., {Basu}, A., {et~al.} 2019, \aap, 632,
  A11

\bibitem[{{Niklas} {et~al.}(1997){Niklas}, {Klein}, \&
  {Wielebinski}}]{niklas97}
{Niklas}, S., {Klein}, U., \& {Wielebinski}, R. 1997, \aap, 322, 19

\bibitem[{{Parker}(1992)}]{parker92}
{Parker}, E.~N. 1992, ApJ, 401, 137

\bibitem[{{Reynolds}(1994)}]{Reynolds94}
{Reynolds}, J.~E. 1994, {ATNF Memo}, {AT/39.3/040}

\bibitem[{{Sanders} {et~al.}(2003){Sanders}, {Mazzarella}, {Kim},
  {et~al.}}]{sanders03}
{Sanders}, D.~B., {Mazzarella}, J.~M., {Kim}, D.~C., {et~al.} 2003, AJ, 126,
  1607

\bibitem[{{Stein} {et~al.}(2020){Stein}, {Dettmar}, {Beck}, {Irwin}, {Wiegert},
  {Miskolczi}, {Wang}, {English}, {Henriksen}, {Radica}, \& {Li}}]{stein20}
{Stein}, Y., {Dettmar}, R.~J., {Beck}, R., {et~al.} 2020, \aap, 639, A111

\bibitem[{{Steinwandel} {et~al.}(2022){Steinwandel}, {Dolag}, {Lesch}, \&
  {Burkert}}]{steinwandel22}
{Steinwandel}, U.~P., {Dolag}, K., {Lesch}, H., \& {Burkert}, A. 2022, \apj,
  924, 26

\bibitem[{{Strickland} \& {Heckman}(2009)}]{strickland09}
{Strickland}, D.~K., \& {Heckman}, T.~M. 2009, \apj, 697, 2030

\bibitem[{{Thompson} \& {Heckman}(2024)}]{thompson24}
{Thompson}, T.~A., \& {Heckman}, T.~M. 2024, \araa, 62, 529

\bibitem[{{Westmeier} {et~al.}(2021){Westmeier}, {Kitaeff}, {Pallot}, {Serra},
  {van der Hulst}, {Jurek}, {Elagali}, {For}, {Kleiner}, {Koribalski},
  {Lee-Waddell}, {Mould}, {Reynolds}, {Rhee}, \& {Staveley-Smith}}]{sofia2}
{Westmeier}, T., {Kitaeff}, S., {Pallot}, D., {et~al.} 2021, \mnras, 506, 3962

\end{thebibliography}
\bibliographystyle{aasjournal}



\end{document}